\documentclass[journal]{IEEEtran_}

\usepackage{cite}
\usepackage[pdftex]{graphicx}

\ifCLASSOPTIONcompsoc
  \usepackage[caption=false,font=normalsize,labelfont=sf,textfont=sf]{subfig}
\else
  \usepackage[caption=false,font=footnotesize]{subfig}
\fi

\usepackage{amsmath}
\usepackage{amssymb}
\usepackage{stackengine}
\usepackage{siunitx}
\usepackage{fancyhdr}
\newcommand{\ve}[1]{\stackunder[1.2pt]{$#1$}{\rule{.8ex}{.075ex}}}
\newcommand{\matriz}[1]{\ve{\ve{#1}}}
\newcommand{\micras}[0]{\si{\micro\metre}}

\begin{document}
\title{Bayesian optimization with improved scalability and derivative information for efficient design of nanophotonic structures}
\author{Xavier~Garcia-Santiago,  Sven~Burger, Carsten~Rockstuhl, 
        and~Philipp-Immanuel~Schneider
\thanks{Xavier Garcia-Santiago is with the Institute
of Theoretical Solid State Physics, Karlsruhe Institute of Technology, 76137 Karlsruhe, Germany and with JCMwave GmbH, Bolivarallee 22, 14050 Berlin, Germany, e-mail: (xavier.garcia-sangiago@kit.edu).}
\thanks{Carsten Rockstuhl is with the Institute
of Theoretical Solid State Physics, Karlsruhe Institute of Technology, 76137 Karlsruhe, Germany, with the Institute of Nanotechnology, Karlsruhe Institute of Technology,  76344 Eggenstein-Leopoldshafen, Germany, and with the Max Planck School of Photonics, Germany.}
\thanks{Sven Burger and Philipp-Immanuel Schneider are with JCMwave GmbH, Bolivarallee 22, 14050 Berlin, Germany and with the Zuse Institute Berlin, Takustra{\ss}e 7, 14195 Berlin, Germany.}
}

\markboth{} 
{Shell \MakeLowercase{\textit{et al.}}: demo}

\maketitle
\thispagestyle{fancy}

\begin{abstract}
We propose the combination of forward shape derivatives and the use of an iterative inversion scheme for Bayesian optimization to find optimal designs of nanophotonic devices. This approach widens the range of applicability of Bayesian optmization to situations where a larger number of iterations is required and where derivative information is available. This was previously impractical because the computational efforts required to identify the next evaluation point in the parameter space became much larger than the actual evaluation of the objective function. We demonstrate an implementation of the method by optimizing a waveguide edge coupler.
\end{abstract}


\section{Introduction}

Finding the optimal design of a device is a fundamental problem in science and engineering. The general solution procedure consists of parametrizing the geometry or materials of the device by a set of variables $\ve{x} = [x_1,x_2,...,x_d]$. By using a numerical solver to determine with an objective function $f_{\rm ob} (\ve{x})$ the performance of devices at different points in the $d$-dimensional parameter space $\mathcal{D}$, one seeks to find an optimal set of parameters $\ve{x}_{\rm opt}$ such that,

\begin{equation}
f_{\rm ob}(\ve{x}_{\rm opt}) \geq f_{\rm ob}(\ve{x}) \, \forall \,\ve{x} \in \mathcal{D}.
\end{equation}

Depending on the parametrization, one can access a different range of possible devices. Usually, the larger the number of parameters the broader the range of designs that can be addressed. However, except for convex problems, the complexity of finding a globally optimal design scales exponentially with the number of parameters. This fact is commonly referred to as the curse of dimensionality\cite{bellman2003dynamic} or the Hughes Effect\cite{hughes1968,hughes_effect}. 

When designing nanophotonic devices, the objective function depends on the electromagnetic field and for its determination one normally needs to solve Maxwell's equations. In general, computing the objective function requires times between some minutes and several hours, depending on the complexity of the problem. This, added to the problem of the exponential scaling of the parameter space, makes finding optimal nanophotonic devices a very hard or even intractable problem. Only for highly symmetric devices, such as devices invariant in one dimension or cylindrically symmetric devices, computing the objective function requires far less times and the global optimization problem can be solved with reasonable effort.

The convergence rate of the optimization problem can often be improved by providing derivatives of the objective function with respect to the design parameters $x_i$. To achieve this, two different methods have been proposed: the direct method\cite{FEM_with_derivatives,hughes2019forward} and the adjoint method\cite{sigmund2003systematic, borel2004topology}. 
The direct method propagates all derivatives with the solution of Maxwell's equations. It is suitable for problems with a low or medium number of parameters, a few dozen at most. The adjoint method determines derivatives by solving additionally an adjoint problem. This additional effort is only worthwhile for high-dimensional parameter spaces. For example, topology optimization typically requires thousands or even millions of degrees of freedom. This makes, however, the problem of finding a global optimum in general intractable\cite{convex}. Nevertheless, designs obtained using local optimization methods show a good performance\cite{molesky2018inverse}.
Here, we consider medium dimensional design spaces, with no more than twenty parameters, where finding a global optimum can still be feasible.  

One highly efficient global optimization algorithm for medium dimensional optimization problems of computationally expensive objective functions is Bayesian optimization~\cite{pelikan1999boa,bull2011convergence,zhang2020segmented, qin2018mantle, qin2019designing}. Bayesian optimization is based on the use of a surrogate stochastic model of the objective function. The stochastic model is used to infer different probabilistic quantities of the objective function with the aim of achieving a faster convergence with respect to the number of evaluations. The stochastic model most frequently used in Bayesian optimization is a Gaussian process~\cite{osborne2009gaussian, jones2001taxonomy}. Gaussian processes are widely used in the field of machine learning~\cite{GP_william, lawrence2004gaussian} for regression and classification tasks. One property to highlight is that they can also incorporate derivative information in a natural manner. In several benchmarks~\cite{schneider2019benchmarking} Bayesian optimization required about an order of magnitude less iterations to find good parameter values when compared to other global optimization methods such as particle swarm optimization or differential evolution. Moreover, the benchmarks showed that providing derivative information can significantly increase the convergence rate.

However, one fundamental disadvantage of Bayesian optimization is its poor scalability with respect to the number of objective function evaluations $N_{\rm ev}$. The time to compute a new sampling point scales with $O(N_{\rm ev}^3)$. Moreover, if also provided with $d$ derivatives of the objective, the sample computation time increases asymptotically by a factor of $\left(1+d\right)^3$.

For very expensive objective functions and a moderate number of iterations the sample computation time is often negligible. 
However, in many cases it can become the major bottleneck of the method.
Specifically, if the simulations are suitable to be parallelized or devices with a high degree of symmetry are considered, e.g., diffractive gratings or cylindrical symmetric devices, the simulation times can be drastically reduced. Often, this allows for solving Maxwell's equations even in a few seconds. The design space can then be sampled very often allowing to explore higher dimensional parameter spaces. In these cases, the $O(\mathrm{N_{\rm ev}^3})$ scaling renders Bayesian optimization impractical. 

There are two main steps during which Bayesian optimization suffers from scalability issues: the inversion of the covariance matrix and the evaluation of the acquisition function~\cite{Schneider2017} at different points. In a previous work~\cite{schneider2019benchmarking}, we proposed a technique to mitigate the problem of the time for evaluating the acquisition function. In this contribution, we propose the use of a matrix update scheme for the inversion of the covariance matrix. With this scheme, the method can scale with $O(N_{\rm ev}^2 (d+1)^2)$ instead of $O(N_{\rm ev}^3 (d+1)^3)$, enabling a significant increase of the number of possible optimization steps in practical applications.

The work starts with a short review of the finite element method with forward shape and material derivatives. Then, the Bayesian optimization method is presented and its scalability problems are demonstrated. After introducing the matrix update approach for addressing the scalability problem, the method is exemplary applied to optimize a waveguide coupler with a minimal spatial footprint as a prototypical problem in integrated nanophotonics.

\section{Numerical procedure}
\subsection{Finite element method with shape derivatives}

The objective function in nanophotonic design problems is typically a function of the electromagnetic field that is a solution of Maxwell's equations. To solve Maxwell's equations, we employ the finite element method\cite{monk1992} (FEM), as it is particularly suitable for handling general shapes of varying complexity. Specifically, we use the commercial solver JCMsuite\cite{JCMsuite_web,pomplun2007Adaptive}. Note however, that the method that we will present in this work can be used with any numerical solver able to calculate the derivatives of an objective function with respect to the design parameters. Other examples are the finite difference time domain method~\cite{hassan2020multilayer}, the beam propagation method~\cite{iguchi2016topology} or the eigenmode expansion method~\cite{burschapers2011sensitivity, petravcek2011bidirectional,murthy1988derivatives}, to name a few.

Within FEM, one solves the weak form of Maxwell's equations approximating the electric field with a discrete set of local basis functions. This calculation translates into an algebraic system
\begin{equation}\label{eq:Maxwells_FEM}
\matriz{A}\ve{e} = \ve{s},
\end{equation}
where $\matriz{A}$ is the system matrix, $\ve{e}$ is the solution of the electric field projected into the basis of the finite element functions, and the right hand side $\ve{s}$ describes source terms like the illumination field. Solving the system \eqref{eq:Maxwells_FEM}, one obtains the solution of the electromagnetic field as
\begin{equation}\label{eq:e_field_FEM}
\ve{e} = \matriz{A}^{-1}\ve{s}.
\end{equation}

In addition to the solution $\ve{e}$, it is also possible to determine the derivatives of the electromagnetic field with respect to the design parameters $x_i$. 

There are two main methods of computing the shape and material derivatives of some merit function $f_{\rm ob}$ that depends on the value of the electromagnetic field: the direct method\cite{FEM_with_derivatives,hughes2019forward} and the adjoint method\cite{veronis2004method}. For both methods the fundamental step is to be able to determine the derivatives of the system matrix $\matriz{A}$ with respect to the design parameters $x_i$, $\frac{d\matriz{A}}{dx_i}$.

In the direct method, the shape or material derivatives of the electric field are obtained as
\begin{equation}
\frac{\mathrm{d}\ve{e}}{\mathrm{d}x_i} = \matriz{A}^{-1}\left(-\frac{\mathrm{d}\matriz{A}}{\mathrm{d}x_i}\ve{e}+\frac{\mathrm{d}\ve{s}}{\mathrm{d}x_i} \right).
\end{equation}

Therefore, once the electric field has been solved, obtaining every derivative $\frac{\mathrm{d}\ve{e}}{\mathrm{d}x_i}$ implies solving the same system of Eq.~\eqref{eq:Maxwells_FEM} with a different right hand side term. Notably, the same convergence properties as for the solution $\ve{e}$ are obtained for $\mathrm{d}\ve{e}/\mathrm{d}x$~\cite{FEM_with_derivatives}. The derivatives $\frac{\mathrm{d}\ve{e}}{\mathrm{d}x_i}$ can then be propagated to compute the derivatives of the objective function $f_{\rm ob}$

\begin{equation}
\frac{\mathrm{d}f_{\rm ob}}{\mathrm{d}x_i} = \frac{\mathrm{d}f_{\rm ob}}{\mathrm{d}\ve{e}^{T}}\cdot\frac{\mathrm{d}\ve{e}}{\mathrm{d}x_i},
\end{equation}
where the vector $\frac{\mathrm{d}f_{\rm ob}}{\mathrm{d}\ve{e}}$ gives the derivatives of the objective function with respect to the values of the solution $\ve{e}$.

The adjoint method determines derivatives by solving aditionally to Eq.~\eqref{eq:Maxwells_FEM} the adjoint problem
\begin{equation}\label{eq:adjoint_system}
\matriz{A}^{\dagger}\ve{\lambda} = \frac{\mathrm{d}f_{\rm ob}}{\mathrm{d}\ve{e}},
\end{equation}
where $\matriz{A}^{\dagger}$ denotes the conjugate transpose of the system matrix $\matriz{A}$.

Once the adjoint system has been solved, obtaining the value of each shape or material derivative involves a matrix-vector multiplication and a vector-vector multiplication 
\begin{equation}
\frac{\mathrm{d}f_{\rm ob}}{\mathrm{d}x_i} = -\ve{\lambda}^{T}\cdot \left(\frac{\mathrm{d}\matriz{A}}{\mathrm{d}x_i}\ve{e}\right).
\end{equation}

In terms of computation times, the main difference between both methods is that the direct method only requires to invert or decompose one system matrix and then to perform two matrix-vector multiplications or one Gaussian elimination and one matrix-vector multiplication per parameter. The adjoint method needs to decompose two different matrices, but once this is done, obtaining the derivatives for each parameter only requires a matrix-vector multiplication and a vector-vector multiplication. This last operation results in an important advantage when the number of parameter derivatives is considerably large as it is, e.g., the case for topology optimization. Then, the adjoint method is especially advantageous because the matrix $\frac{\mathrm{d}\matriz{A}}{\mathrm{d}x_i}$ is mainly composed of zeros, as each parameter $x_i$ represents usually the permittivity value of a small pixel. The matrix-vector operation can then be neglected and the time for obtaining each derivative parameter is determined just by the vector-vector multiplication.

For the opposite reason, in this work the direct method is used. The goal is to optimize problems with around 20 parameters at maximum. For this case, solving the additional adjoint system typically imposes a severe overhead with respect to the direct method and the non zero entries of $\frac{\mathrm{d}\matriz{A}}{\mathrm{d}x_i}$ are, in general, considerably larger than for topology optimization problems. Moreover, using forward derivatives allows to work with objective functions of higher complexity, as the derivatives can be propagated for the objective function in a simpler way using automatic differentiation \cite{bartholomew2000automatic}. Reference~\cite{kepler2010sensitivity} provides a detailed analysis and comparison between both the forward and the adjoint methods.

The FEM solver used in our calculations allows to compute the shape and material derivatives of different quantities derived from the fields, as for example the amplitudes of its angular spectrum representation. 
To compute the shape derivatives, we only need to specify 
the partial shape derivatives of the nanophotonic structure, i.e., how the boundaries of the different elements of the structure change with respect to an infinitesimal change in the design variables. Then the solver automatically obtains the derivatives of the mesh elements and propagates them until it obtains the derivatives of the final quantity of interest.

\subsection{Bayesian optimization with Gaussian processes}

A Gaussian process $\mathcal{GP}(m,k)$ is a stochastic model, which yields a probabilistic distribution over functions (Ref.~\cite{GP_william}, Chapter 2.2). We use it as a surrogate model of the computationally expensive objective function $f_{\rm ob}(\ve{x})$.
A Gaussian process (GP) is characterized by its mean $m\left(\ve{x}\right)$ and covariance functions $k\left(\ve{x},\ve{x'}\right)$. We parametrize the mean function as,

\begin{equation}\label{eq:mean_function}
m\left(\ve{x}\right) = m_0,
\end{equation}
and the covariance function with the Mat\'ern 5/2\cite{stein2012interpolation} covariance function
\begin{equation}
\label{eq:matern_kernel}
    k\left(\ve{x},\ve{x'}\right) = \sigma_0^2\left(1+\sqrt{5}r+\frac{5}{3}r^2\right)e^{-\sqrt{5}r},
\end{equation}
where
\begin{equation}\label{eq:length_scales}
    r^2=\sum_{i=1}^d \frac{(x_i-x_i')^2}{l_i^2}.
\end{equation}

The hyperparameters $m_0, \sigma_0, l_1, \cdots,l_d$ are chosen to maximize the log-likelihood of all observations of the function $f_{\rm ob}(\ve{x})$~\cite{GP_william}. The hyperparameter optimization is computationally expensive. Therefore, we perform it only if required~\cite{garcia2018shape} and stop their optimization after 300 observations.

The probabilistic behaviour of a function $f\left(\ve{x}\right)$ modelled as a GP at any collection of points $\matriz{X}_{1:N_{\rm ev}} = \left\{ \ve{x}_1,...,\ve{x}_{N_{\rm ev}}\right\}$ is described by a multivariate Gaussian distribution over these points
\begin{equation}
\ve{f}_{1:N_{\rm ev}} \sim \mathcal{N}\left(\ve{m}_{1:N_{\rm ev}},\matriz{K}_{1:N_{\rm ev}}\right),
\end{equation}
where $\ve{f}_{1:N_{\rm ev}}$ is the random vector of function values $[f(\ve{x}_1),\cdots,f(\ve{x}_{N_{\rm ev}})]^T$, $\ve{m}_{1:N_{\rm ev}}$ the vector of mean values $[m(\ve{x}_1),\cdots,m(\ve{x}_{N_{\rm ev}})]^T$, and $\matriz{K}_{1:N_{\rm ev}}$ the covariance matrix, where the $i$-$j$ element is determined by $k(\ve{x}_i,\ve{x}_j)$.

Let us assume that $f_{\rm ob}(\ve{x})$ has been evaluated at some points $\matriz{X}_{\rm ev} = \left[ \ve{x}_{{\rm ev}1},...,\ve{x}_{{\rm ev}N}\right]$ and let $\matriz{X}_{*} = \left[ \ve{x}_{*1},...,\ve{x}_{*M}\right]$ denote another set of points. The prior joint distribution for the two sets can be calculated with the mean $m(\ve{x})$ and covariance $k(\ve{x},\ve{x'})$ functions (Ref.~\cite{GP_william}, Eq. 2.18)

\begin{equation}\label{eq:prior_GP}
\begin{bmatrix}\ve{f}_{\rm ev}\\ \ve{f}_{*}\end{bmatrix} \sim \mathcal{N}\left(\begin{bmatrix} \ve{m}_{\rm ev}\\\ve{m}_{*} \end{bmatrix},\begin{bmatrix}\matriz{K}_{\rm ev} & \matriz{K}_{\rm ev,*}\\ \matriz{K}_{\rm ev,*}^{T} & \matriz{K}_{*}\end{bmatrix}\right),
\end{equation}
where $\matriz{K}_{\rm ev}$ is the covariance matrix for the set $\matriz{X}_{\rm ev}$, $\matriz{K}_{*}$ the covariance matrix for the set $\matriz{X}_{*}$, and $\matriz{K}_{\rm ev,*}$ the crossvariance between the two sets. 

Using Bayes's theorem~\cite{bayes1763lii}, one can determine the posterior probability distribution over function values $\ve{f_{*}}$ at $\matriz{X}_{*}$ given the observations $(\matriz{X}_{\rm ev},\ve{f}_{\rm ev})$. The posterior distribution
\begin{equation}
\ve{f_{*}}|\matriz{X}_{*},\matriz{X}_{\rm ev},\ve{f}_{\rm ev} \sim \mathcal{N}\left(\ve{m}_p,\matriz{K}_p\right)
\end{equation}
is again a multivariate Gaussian distribution with the posterior mean vector $\ve{m}_p$ and covariance matrix $\matriz{K}_p$
\begin{align} 
\ve{m}_p &= \ve{m}_{*} + \matriz{K}_{*,\rm ev}\matriz{K}^{-1}_{\rm ev}\left(\ve{f}_{\rm ev}-\ve{m}_{\rm ev}\right), \label{eq:inference_mean_no_derivatives}\\ 
\matriz{K}_p &= \matriz{K}_{*} - \matriz{K}_{*,\rm ev}\matriz{K}_{\rm ev}^{-1}\matriz{K}_{*,\rm ev}^{T}. \label{eq:inference_sigma_no_derivatives}
\end{align}

\begin{figure}[htb]
\subfloat{
\includegraphics[width=\linewidth]{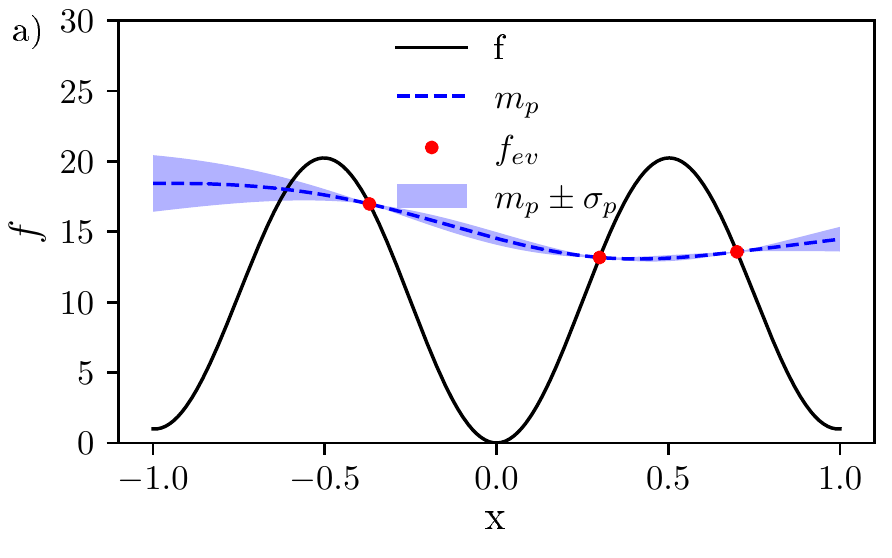}
}
\hfill
\subfloat{
\includegraphics[width=\linewidth]{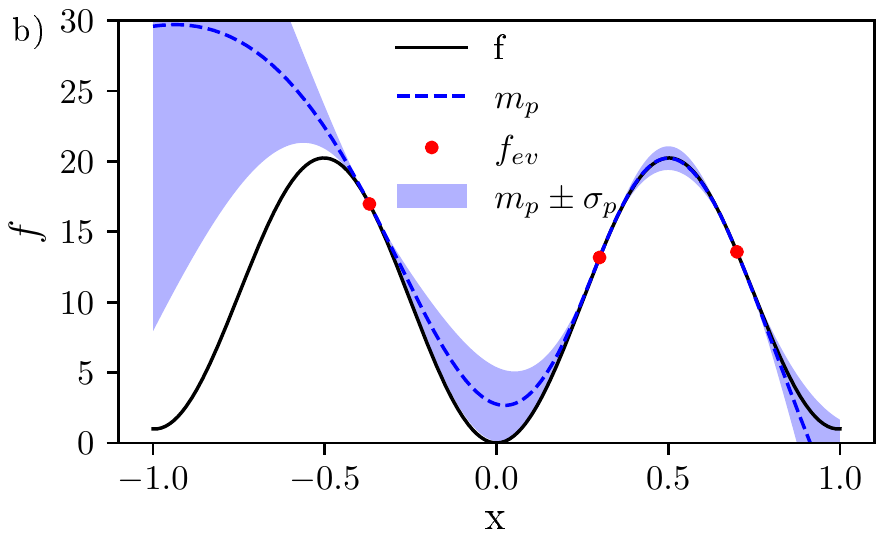}
}
\caption{Example of Gaussian processes regression. The figure shows the values of the posterior mean and standard deviation, Eqs. \eqref{eq:inference_mean}-\eqref{eq:inference_sigma}, from three observations $f_{\rm ev}$ of a function $f(x)$. Top: GP regression using the information of the evaluations of the function $f$. Bottom: The derivative of the function $f$ with respect to the input variable $x$ has been also included into the GP regression and the hyperparameter estimation. The standard deviation has larger values for the regression including derivative observations, but this fact results in a better estimation of the real behavior of function $f$.}
\label{fig:GP_with_derivatives}
\end{figure}

The above two equations give a probabilistic distribution for the values of the objective function $f_{\rm ob}(\ve{x})$ at any point $\ve{x_{*}}$ where we do not know its true value, based on the information we have obtained from the evaluations already computed. The diagonal elements of $\matriz{K}_p$ contain the variance $\sigma_p^2$ of the different points of $\matriz{X}_{*}$. Figure~\ref{fig:GP_with_derivatives}a shows the values of $m_p$ and $\sigma_p$ for a function $f(x)$ with three known function values.

The Bayesian optimization method uses the statistical information provided by the posterior GP to determine promising parameter sets and to sequentially converge to the optimum value of $f_{\rm ob}$. There are different strategies for how to use the information provided by the GP~\cite{Schneider2017}. One commonly used strategy, which is also used in this work, consists in finding the point of maximum expected improvement~\cite{GP_william}.

It addition to the function values itself, it is also possible to include derivative observations to obtain more accurate GP predictions, as, e.g., described in Ref.~\cite{papoulis2002probability} chapter 10 or Ref.~ \cite{GP_william} chapter 9.4. The use of derivative information for regression with Gaussian processes was already used to model dynamic systems~\cite{solak2003derivative} and also applied to Bayesian optimization~\cite{wu2017BOgradients,SSP_GP, schneider2019benchmarking}.
The covariance function of the joint process of objective values and first derivatives
\begin{equation}
\begin{bmatrix} f(\ve{x}) \\ \nabla f(\ve{x}) \end{bmatrix} \sim \mathcal{GP}\left( 0, \matriz{K_{f,\nabla f}} \right)
\end{equation}
can be split into the below four blocks~\cite{SSP_GP}, 
\begin{equation}\label{eq:k_with_gradients}
\matriz{K_{f,\nabla f}}(\ve{x},\ve{x'}) = \begin{bmatrix} k_{[f,f]}(\ve{x},\ve{x'}) & k_{[f,\nabla f]}(\ve{x},\ve{x'}) \\ k_{[\nabla f,f]}(\ve{x},\ve{x'}) & k_{[\nabla f,\nabla f]}(\ve{x},\ve{x'}) \end{bmatrix}
\end{equation}
with the covariance functions $k_{[f,f]}(\ve{x},\ve{x'})$, $k_{[f,\nabla f]}(\ve{x},\ve{x'})$, $k_{[\nabla f,f]}(\ve{x},\ve{x'})$, and $k_{[\nabla f,\nabla f]}(\ve{x},\ve{x'})$ being
\begin{align}
k_{[f,f]}(\ve{x},\ve{x'}) &= cov(f(\ve{x}),f(\ve{x'})) = k(\ve{x},\ve{x'}), \label{eq:covariance_with_derivatives_1}\\
k_{[f,\nabla f]}(\ve{x},\ve{x'}) &= cov(f(\ve{x}),\nabla f(\ve{x'})) = \nabla _{\ve{x'}} k(\ve{x},\ve{x'}), \\
k_{[\nabla f,f]}(\ve{x},\ve{x'}) &= cov(\nabla f(\ve{x}),f(\ve{x'})) = \nabla _{\ve{x}}k(\ve{x},\ve{x'}), \\
k_{[\nabla f,\nabla f]}(\ve{x},\ve{x'}) &= cov(\nabla f(\ve{x}),\nabla f(\ve{x'})) = \nabla _{\ve{x}}\nabla _{\ve{x'}}k(\ve{x},\ve{x'}). \label{eq:covariance_with_derivatives_4}
\end{align}

The same inference rules of the posterior distributions as in Eqs.~\eqref{eq:inference_mean_no_derivatives}-\eqref{eq:inference_sigma_no_derivatives} also apply for the larger process. That is, the posterior distribution over $\ve{f_*}$ can be obtained as~\cite{SSP_GP}
\begin{align} \label{eq:inference_mean}
\ve{m}_p(\ve{x}^*) &= \matriz{K}_{*,{\rm ev}}{\matriz{K}_{\rm ev}^{\nabla f}}^{-1}\left[\ve{f}_{\rm ev}^T,\nabla \ve{f}_{\rm ev}^T\right]^T \\ 
\matriz{K}_p(\ve{x}^*) &= \matriz{K}_{*} - \matriz{K}_{*,{\rm ev}}^{\nabla}{\matriz{K}_{\rm ev}^{\nabla f}}^{-1}{\matriz{K}_{*,{\rm ev}}^{\nabla f}}^T, \label{eq:inference_sigma}
\end{align}
where $\matriz{K}_{*}$ is the same covariance matrix as in Eq.~\eqref{eq:prior_GP}, $\matriz{K}_{\rm ev}^{\nabla f}$ is the covariance matrix for the set $\matriz{X}_{\rm ev}$ obtained evaluating $\matriz{K}_{f,\nabla f}(\ve{x},\ve{x'})$, and $\matriz{K}_{*,{\rm ev}}^{\nabla f}$ is the crossvariance between the two sets $\matriz{X}_{*}$ and $\matriz{X}_{\rm ev}$ obtained also using Eqs.~\eqref{eq:covariance_with_derivatives_1}-\eqref{eq:covariance_with_derivatives_4}. For simplification of notation, from this point we will refer with the term $\matriz{K}_{\rm ev}$ to the covariance matrix of the GP for both the cases with and without derivative information. The rank of the matrix is determined by the number of derivative and non-derivative observations $N_{\rm obs} = N_{\rm ev}\cdot (d+1)$.

In Fig.~\ref{fig:GP_with_derivatives} we can see how adding derivative observations increases the accuracy of the GP representation of the function in a vicinity of the evaluation points. The extra information also provides a better estimation of the hyperparameters in Eqs.~\eqref{eq:mean_function} and \eqref{eq:matern_kernel} and thus of the uncertainty $\sigma_p$. Due to the more accurate GP predictions, derivative information can significantly improve the convergence rate of Bayesian optimization.

On the other hand, as can be seen from Eq.~\eqref{eq:k_with_gradients}, including the derivative information of $d$ parameters implies to introduce $d$ extra entries in the covariance matrix after every evaluation. This increases the effort to compute the right-hand side of Eqs.~\eqref{eq:inference_mean}-\eqref{eq:inference_sigma}.

The covariance matrix $\matriz{K_{\rm ev}}$ becomes typically ill-conditioned during the course of the optimization. Therefore, the computation and application of its inverse is avoided. Since $\matriz{K_{\rm ev}}$ is symmetric and positive definite, a numerically more stable approach is to determine its Cholesky decomposition $\matriz{L}_{K_{\rm ev}}$ in $O(N_{\rm obs}^3)$ steps and solve the equation 
\begin{equation}
\matriz{K}_{\rm ev}\cdot\ve{b} = \matriz{L}_{K_{\rm ev}}\cdot\matriz{L}_{K_{\rm ev}}^T\cdot\ve{b} = \ve{x}
\end{equation}
by forward and backward substitution in $O(N_{\rm obs}^2)$ steps. Other approaches~\cite{GPregularization, foster2009stable, SSP_GP} that address the ill-condition of the covariance matrix are discussed in the Appendix \ref{sec:appendix_stability}.

Figure~\ref{fig:inversion_times_n_ev} shows the time to determine the Cholesky decomposition both with and without derivative information for $d=10$ parameters. For the case of nanophotonic structures that are solvable on the time scale of seconds, already after thousand simulations decomposing the covariance matrix requires around the same time as solving Maxwell's equations when including the derivative information. For ten thousand simulations it would require around twenty minutes, rendering Bayesian optimization impractical.

\begin{figure}[htb]
\centering
\includegraphics[width=\linewidth]{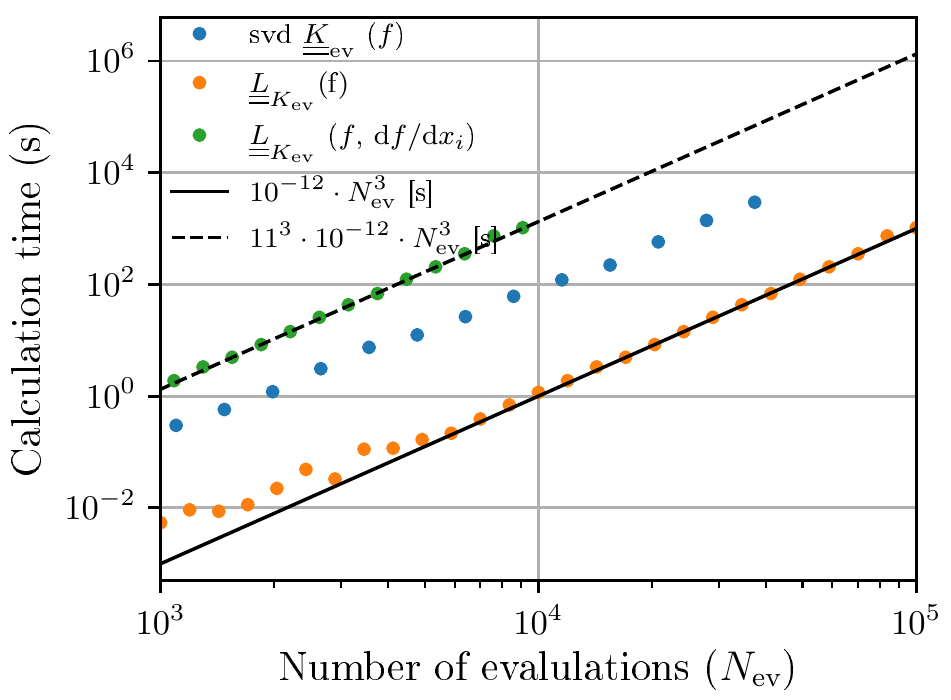}
\caption{Time needed to perform the Cholesky decomposition of the covariance matrix $\matriz{K_{\rm ev}}$ as a function of the number of evaluations of a function $f: \mathbb{R}^{10} \rightarrow \mathbb{R}$ (orange and green points). For comparison, also the time to perform a singular value decomposition (SVD) is included (blue points). The orange and blue points correspond to the case of observing only the function value $f$ at each evaluation. The green points correspond to the case of observing also the derivatives $\mathrm{d}f/\mathrm{d}x_i$ with respect to all ten parameters. The lines show the asymptotic behavior for the calculation of $\matriz{L_{K_{\rm ev}}}$. The calculations were done on a cluster node with 250 GB of memory, due to the memory requirements of Gaussian processes for data sets containing a large number of observations.}
\label{fig:inversion_times_n_ev}
\end{figure}

In order to still be able to treat these optimization problems using Bayesian optimization, we propose the use of a matrix update method to compute the Cholesky decomposition of the covariance matrix. 
The main idea of the approach is to take advantage of the fact that at each iteration, a large block of the covariance matrix has been already decomposed in the previous iteration.
A block inversion or decomposition of a matrix is not numerically faster than the standard algorithms used for this task, unless one has already pre-computed the decomposition of a large block of the matrix. For this reason, a matrix update method for the GP regression is only of advantage if the GP is iteratively updated and predictions are required at each iteration as in the case of Bayesian optimization. 

\begin{figure}[htb]
\centering
\includegraphics[width=\linewidth]{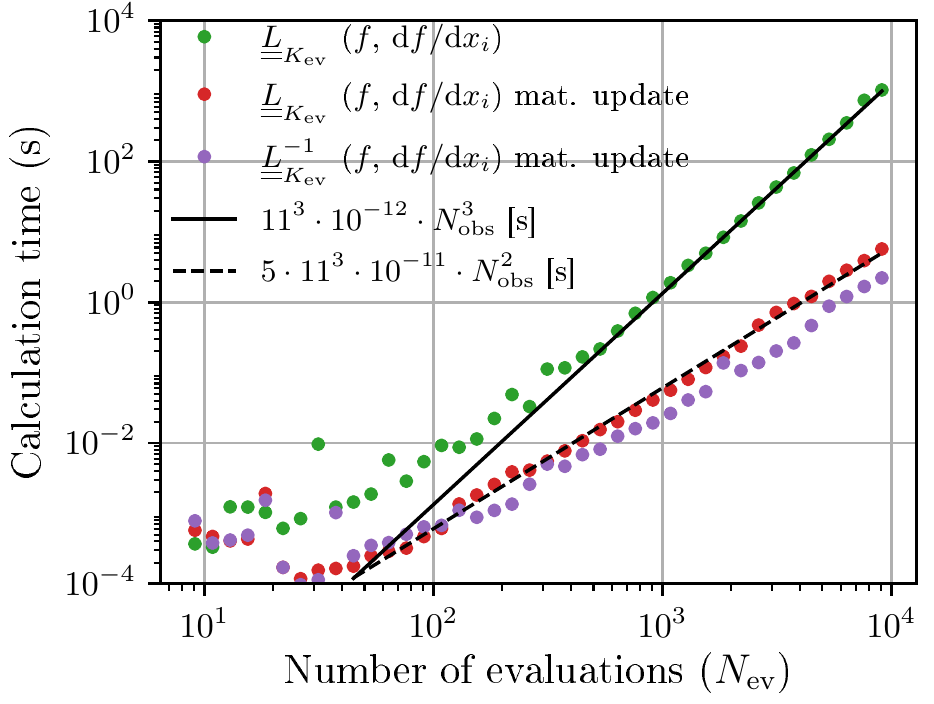}
\caption{Time needed to perform the decomposition of the covariance matrix $\matriz{K}_{\rm ev}$ as a function of the number of evaluations ($N_{\mathrm{ev}}$). Each evaluation is composed of the observation of the function value, $f$, and of its derivatives, $\mathrm{d}f/\mathrm{d}x_i$, with respect to ten parameters. Three different operations are considered, the calculations of the Cholesky decomposition (green points), $\matriz{L}_{K_{\rm ev}}$, the iterative calculation of $\matriz{L}_{K_{\rm ev}}$ (red points) and the iterative calculation of $\matriz{L}_{K_{\rm ev}}^{-1}$ (purple points). Lines mark the asymptotic behavior of the computation times.}
\label{fig:inversion_times}
\end{figure}

The numerical procedure to update $\matriz{L}_{K_{\rm ev}}$ is shown in appendix~\ref{sec:appendix_matrix_update}. Computing $\matriz{L}_{K_{\rm ev}}$ iteratively using the previous $\matriz{L}_{K_{\rm ev}}$ reduces the numerical effort from $O(N_{\rm obs}^3)$ to $O(N_{\rm obs}^2)$, as shown in Fig.~\ref{fig:inversion_times}. The numerical implementation is based on the C implementation of the BLAS\cite{lawson1979basic} and LAPACK\cite{anderson1990lapack} libraries.

Additionally to the iterative construction of the Cholesky decomposition, it is also possible to determine the inverse of the Cholesky decomposition $\matriz{L}_{K_{\rm ev}}^{-1}$ in an iterative way (see Appendix~\ref{sec:appendix_matrix_update}). As it holds that $\matriz{K}_{\rm ev}^{-1} = \matriz{L}_{K_{\rm ev}}^{-T}\cdot\matriz{L}_{K_{\rm ev}}^{-1}$, it is possible to directly solve the inference of Eqs.~\eqref{eq:inference_mean}-\eqref{eq:inference_sigma}.

With this second iterative approach one can halve the inference time as only a matrix vector multiplication has to be performed instead of a forward and backward substitution. However, we found that the matrix update of this second scheme is numerically much more unstable, as shown in Appendix~\ref{sec:appendix_stability}.

We like to note, that the iterative update of the Cholesky matrix is not applicable if the GP hyperparameters are updated at each iteration. In this case, the full covariance matrix is updated and a full Cholesky decomposition is required. On the other hand, after some hundred observations the hyperparameter optimization becomes numerically extremely expensive since a series of Cholesky decompositions has to be performed. Moreover, practicable values of the hyperparameters can usually be estimated based on fewer observations.

\section{Waveguide coupler for inter-chip optical communication}

\begin{figure}[h!]
\centering
    \includegraphics[width=0.7\linewidth]{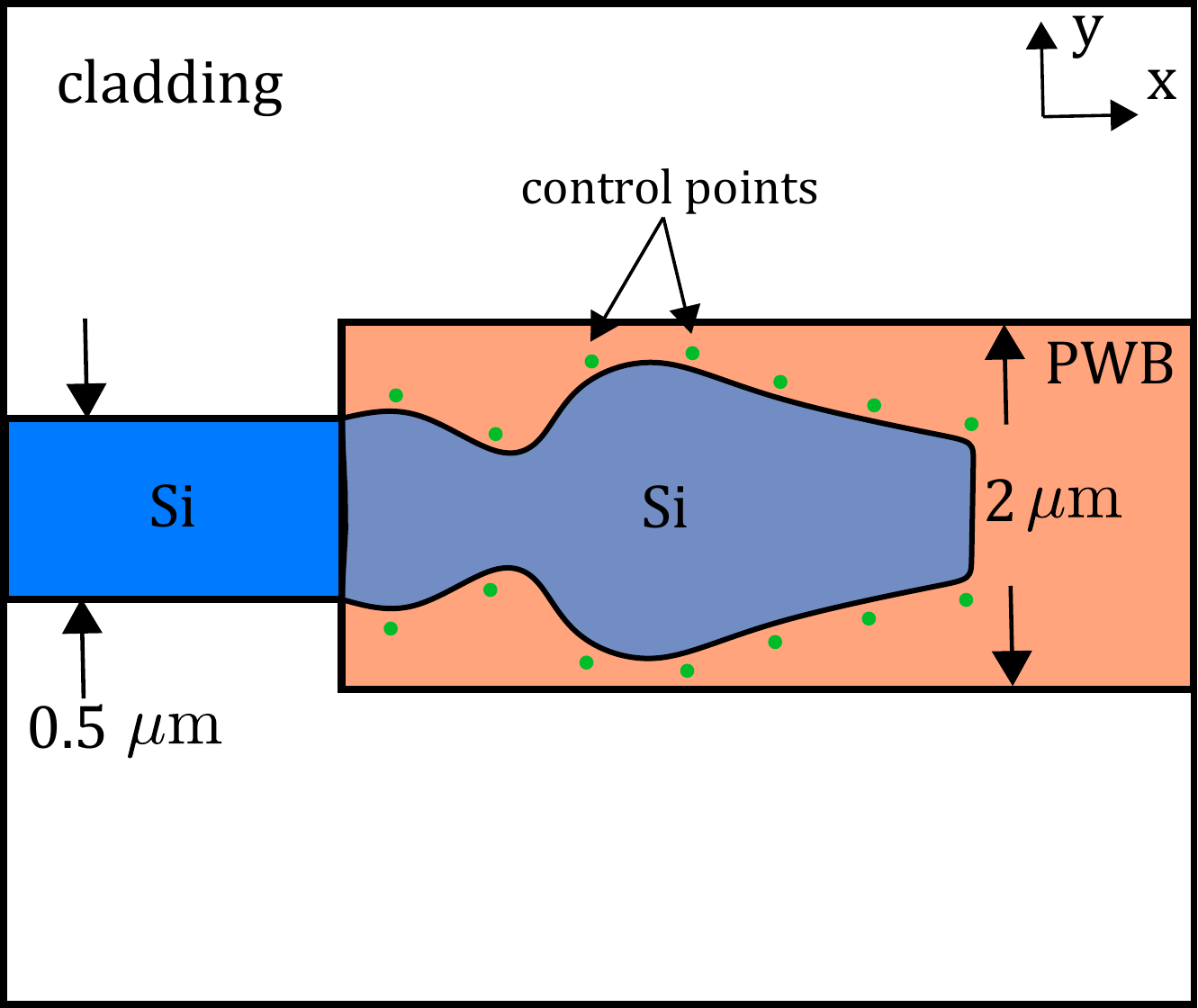}
\caption{Schematic of the waveguide coupler between the silicon waveguide and the photonic wire bond waveguide. The shape of the coupler is described by a NURBS curve. The curve has a number $d$ of control points, equispaced along the x direction. The total length of the coupler is 3 $\mathrm{\mu}m$.}
\label{fig:schematic_coupler}
\end{figure}

To demonstrate the applicability of our approach to design nanophotonic structures, we apply it to optimize a freeform waveguide coupler with a minimal spatial footprint as shown in the schematic of Fig.~\ref{fig:schematic_coupler}. The waveguide coupler is part of a photonic inter-chip communication architecture \cite{Aleksandar}. It consists of a silicon based integrated photonic chip that shall be connected to a different photonic chip by means of a photonic wire bond (PWB). The PWB is a freeform dielectric waveguide that can be written with 3D laser nanolithography to follow an arbitrary path. To keep the connection ideally lossless, the PWB shall have an optimal trajectory~\cite{Fernando,hammer2015full} but also the connection between the silicon waveguide and the PWB should be optimized. We concentrate here on the second challenge and our objective is to design a coupler with a minimal spatial footprint that couples the fundamental mode of a silicon waveguide into the fundamental mode of a PWB at a central wavelength of $\lambda _c$ = 1550 nm. 

The silicon waveguide has standard dimensions for an integrated photonic circuit, i.e. we consider a width of 500 nm and a height of 220 nm. The PWB has a width of 2 \micras{} and a height of 1.8 \micras{}. Its refractive index at $\lambda_c$ is 1.53. Both waveguides are placed on top of a Silica substrate, $n_{\rm SiO_2}$ = 1.44, and are surrounded by a cladding with a refractive index of 1.36.

Adiabatic tapers are the most standard structure used for edge couplers~\cite{smith1996reduced,ramadan1998adiabatic} although different structures have been proposed~\cite{wang2017cantilever,teng2019trident,hammer2019oblique,mu2020edge}. To achieve low coupling losses and a compact device design, many different structures have been proposed, such as tapers with smooth non uniform profiles~\cite{agrebi2012coupling,hettrick2004experimental,liu2017design} or with more complex shapes~\cite{sethi2017ultra}. Generally, the more compact the device, the more complex needs to be its shape and also the smaller is its bandwidth. Many different ultra-compact devices have been achieved using topology optimization~\cite{teng2020miniaturized,liu2018adiabatic}. However, the fabrication of these devices is complex.
An alternative is to design a freeform taper. This approach has been used to design similar silicon photonic elements, such as power splitters ~\cite{sideris2019ultrafast,lalau2013adjoint}. Here, we follow this approach to design a waveguide coupler with a length of 3 \micras{}. The shape of the coupler is parametrized with a mirror symmetric NURBS curve, with a series of $d$ control points. The control points are equispaced in the x dimension and they are allowed to vary in the y dimension. The width of the coupler is constrained to be larger than 140 nm and smaller than 1800 nm. These dimensions are compatible with fabrication processes using e-beam lithography.

The simulation of the full 3D structure is computationally very demanding. In order to reduce the computation time and to be able to assess how the coupling efficiency improves when more complex shapes are considered, i.e., when more control points are included, we reduce the three dimensional problem to an effective two dimensional approximation. In the two dimensional simulations, the refractive indices of the silicon waveguide and PWB are replaced by the effective refractive indices of their fundamental modes~\cite{chung1990assessment} computed with a FEM mode solver \cite{JCMsuite_web}. The refractive index used for the cladding corresponds to the one of the cladding of the 3D model, 1.36. The layout of the considered model corresponds to the x-y top plane shown in Fig.~\ref{fig:schematic_coupler}. 

The objective function used for the optimization is the mode overlap of the scattered field, $\ve{E}_{{\rm scat}}$, and the fundamental mode of the PWB, $\ve{E}_{{\rm PWB},0}$ along the cross section $S$, 

\begin{equation}\label{eq:f_ob_waveguide_coupler}
f_{\rm ob} = \left|\int{\left(\ve{E}_{{\rm scat}}\times\ve{H}_{{\rm PWB},0}\right)\cdot\ve{dS}}\right|^2.
\end{equation}

The silicon waveguide is excited with its fundamental mode $\ve{E_{{\rm Si},0}}$. For the simulations of the scattered field, we use a FEM with a polynomial order of 3 and a mesh size of $\mathrm{\lambda}$/10, where $\mathrm{\lambda}$ is the wavelength inside the material. Each simulation requires around 15 seconds to obtain both the coupling efficiency and its shape derivatives with respect to the control points of the NURBS curve. 

We have performed optimizations for waveguide couplers parameterized with a different number of control points $d$ equal to  4, 10, 15, and 20. To test the convergence rate of the optimizations, every optimization is repeated four times. 

\begin{figure}[htb]
\centering
\includegraphics[width=\linewidth]{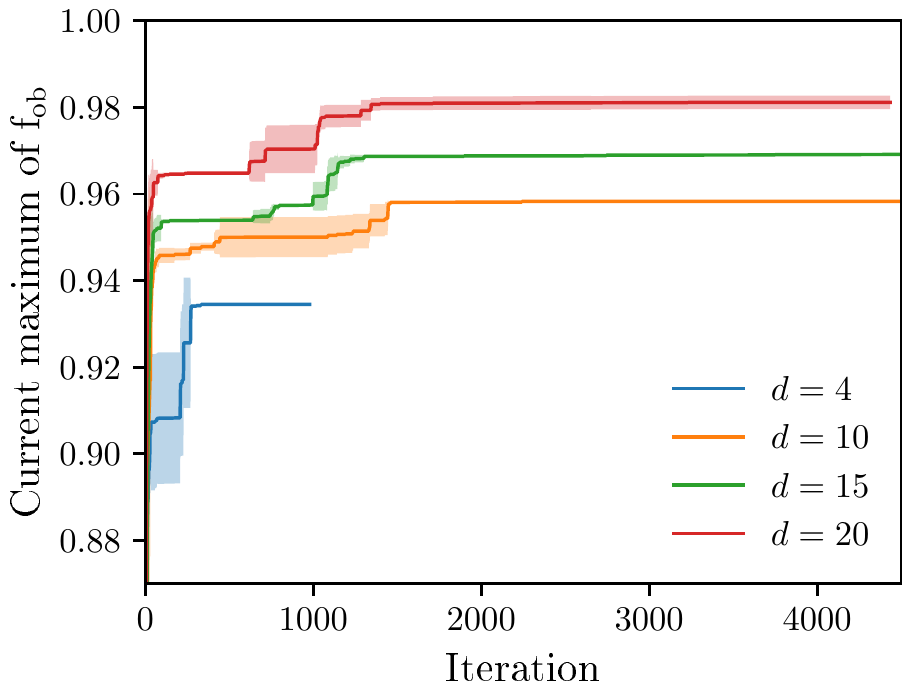}
\caption{Current optimal value of $f_{{\rm ob}}$ during the optimization process of the waveguide coupler. The results are shown for waveguide couplers parametrized with a different number $d$ of control points. The line shown for each $d$ is the mean value obtained over four independent optimization runs. The shadowed area represents the region within one standard deviation.The optimizations for $d$ equals to 4 were stopped after one thousand simulations, as the maximum values of the expected improvement were already lower than $10^{-10}$.}
\label{fig:optimization_results}
\end{figure}

The results of the optimizations are shown in  Fig.~\ref{fig:optimization_results}. They show the average convergence rate of the four optimizations and its standard deviation. We can see that the optimizations for different $d$ converge to different values. The efficiency of the coupler increases with the value of $d$. This result can be expected, as the shape of the devices represented with fewer control points are contained in the design spaces of the larger parametrizations. However, increasing the number of parameters $d$ could also slow down the convergence rate due to the exponentially growing search space. Each optimization exploits the data of $N_{\mathrm{ev}}\cdot(d+1)$ observations, being $N_{\mathrm{ev}}$ the total number of iterations used. One can measure the percentage of the explored volume of the parameter space as the ratio between the number of observations used and the number of $d$-dimensional length scale blocks with side lengths $l_i$ that composes the parameter space. The length scales are defined in Eqn.~\eqref{eq:length_scales}. This measure gives values of 20.5, 10.2, 1.0, and 0.02 percent for the optimizations with $d$ equals to 4, 10, 15, and 20 respectively.   

\begin{figure}[htb]
\centering
\includegraphics[width=\linewidth]{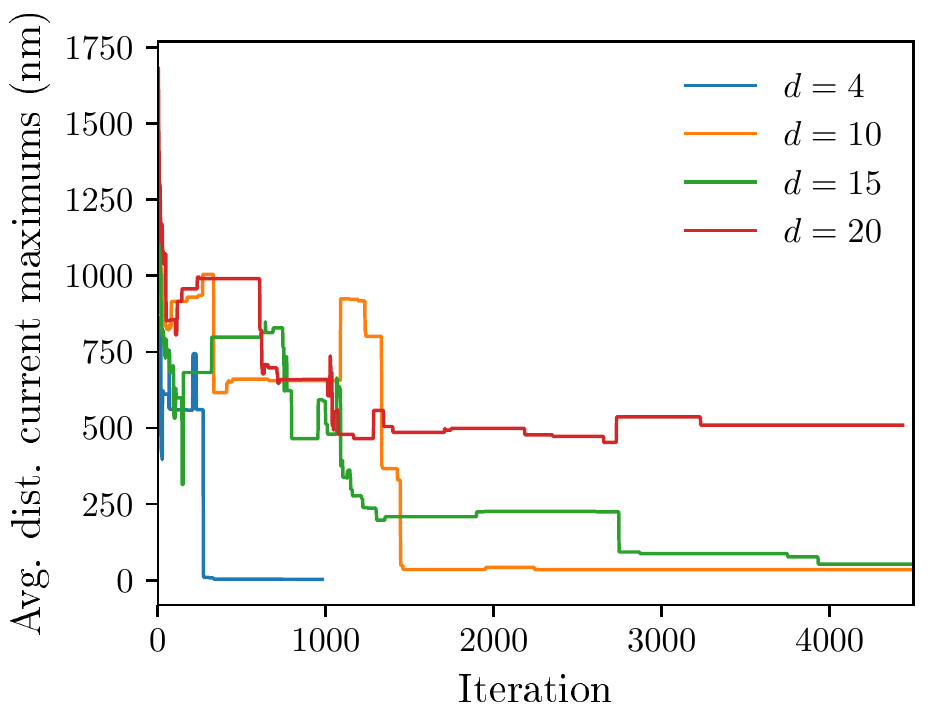}
\caption{Mean value of the distance between the optimal points obtained in different optimization runs. The results are shown for the optimization of waveguide couplers parametrized with a different number $d$ of control points.}
\label{fig:optimization_distances}
\end{figure}

Indeed, for a growing number of parameters it takes increasingly longer to locate the global maximum. This can be seen in Fig.~\ref{fig:optimization_distances}, where the average distance to the best parameter configuration of all four optimization runs are shown as a function of the iteration number. For the optimizations with 15 parameters, around four thousand simulations are required for all four optimization runs to approach the same optimum. For the optimizations with 20 parameters, the optimizations did not converge to the same design even after 4500 simulations.   
However, the standard deviation of the coupling efficiency converged to 0.001 already after one thousand simulations. This result indicates that for these high dimensional spaces there are multiple designs showing a similar performance. Similar results are also frequently obtained in topology optimizations that use the adjoint method. 

Altogether, using a larger number of parameters $d$ for the optimization seems to be favourable. Although it takes longer to reach a global maximum, the coupling efficiencies are on average better already after a small number of iterations for larger $d$. This shows, that the approach scales well for increasing the number of design parameters. The use of shape derivatives probably plays a crucial role in this behavior. 

\begin{figure}[htb]
\centering
\includegraphics[width=\linewidth]{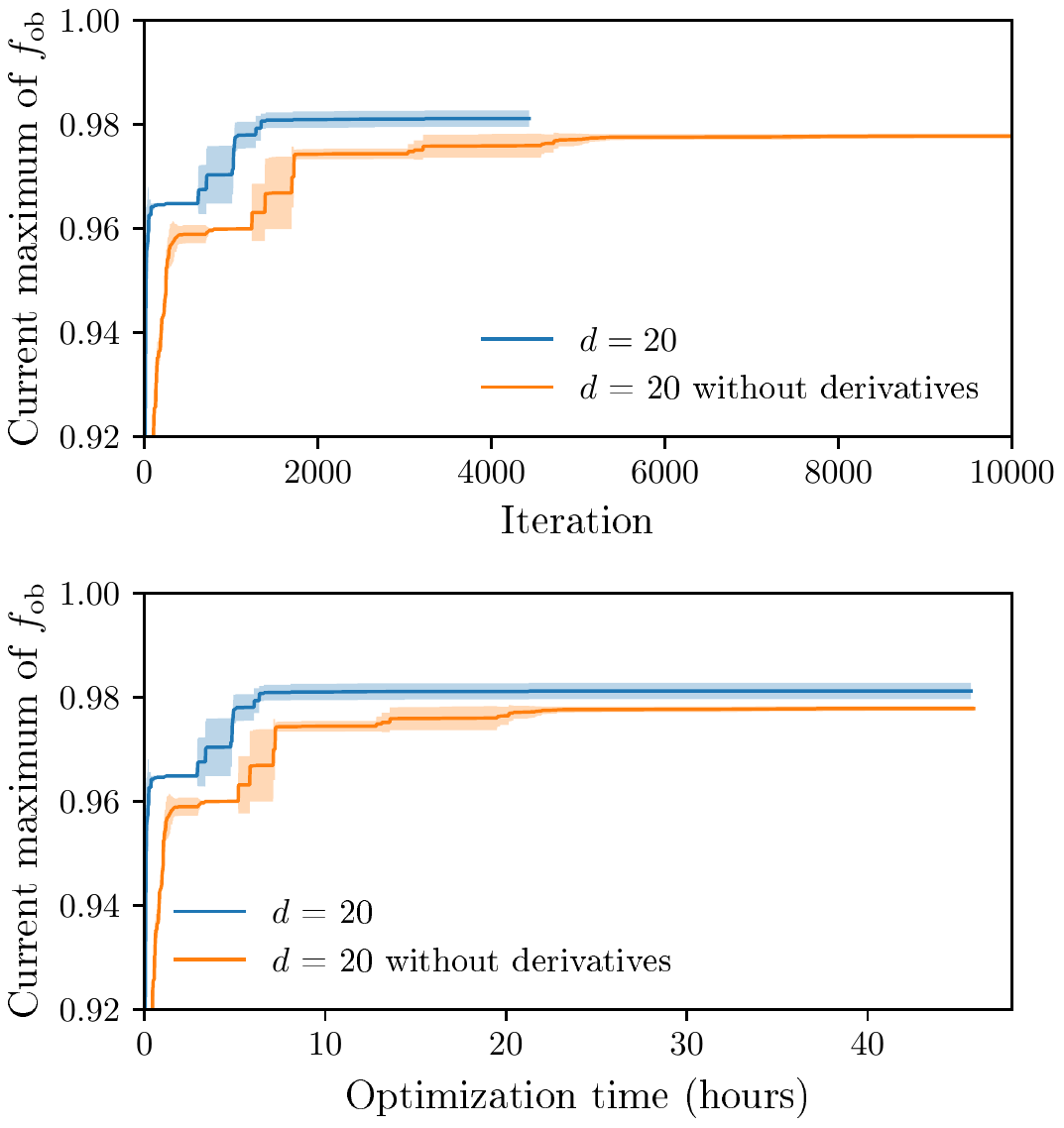}
\caption{Comparison of the performance between two Bayesian optimization algorithms with and without derivative information. The comparison is done for the optimization of the waveguide coupler with $d$ = 20. Top: Current optimum of the coupling efficiency with respect to the number of iterations, FEM simulations, performed by the optimization algorithm. Bottom: Optimum of the coupling efficiency with respect to the run time of the algorithm.}
\label{fig:benchmark_bayesian}
\end{figure}

To further analyze the role played by the shape derivatives in the convergence of the algorithm, we compare the results obtained with the proposed algorithm with those obtained using the same Bayesian optimization algorithm without including derivative information. The comparison is done for the coupler parametrized with 20 parameters, $d$ = 20. The number of iterations for the optimizations without derivatives are set such that the total time used for both optimization methods is approximately the same. The results are shown in Fig.~\ref{fig:benchmark_bayesian}. As one can see, the algorithm that used derivative information obtained slightly better results and it required around one third of the time to obtain coupling efficiencies as the best one obtained by the algorithm without derivative information.

Let us note that two thousand evaluations with twenty parameters using shape derivatives implies already to have a covariance matrix of size 40,000 x 40,000. Decomposing this matrix without using the matrix update scheme would already require computational times around one and a half minutes, compared with the 15 seconds needed for the FEM simulation. With the iterative approach, on the other hand, the decomposition for this matrix size takes only 1 second. Without the use of the iterative approach, the algorithm that exploits derivative information would require 276 hours instead of 47 hours to perform the 4500 iterations. The determination of a parameter set with large expected improvement for the optimization case with $d$ = 20 parameters took around 12 seconds. This number is the average over all the iterations. Let us note also that with the proposed approach the memory still scales with $\mathcal{O}(N_{\rm obs}^2)$. The representation of the covariance matrix and its Cholesky decomposition requires to store $2 \cdot N_{\mathrm{obs}}^2$ double precision floating point numbers, each taking 8 bytes of memory. The optimization of the coupler with 20 parameters needed 120 gigabytes.

\begin{figure}[htb]
\centering
\includegraphics[width=\linewidth]{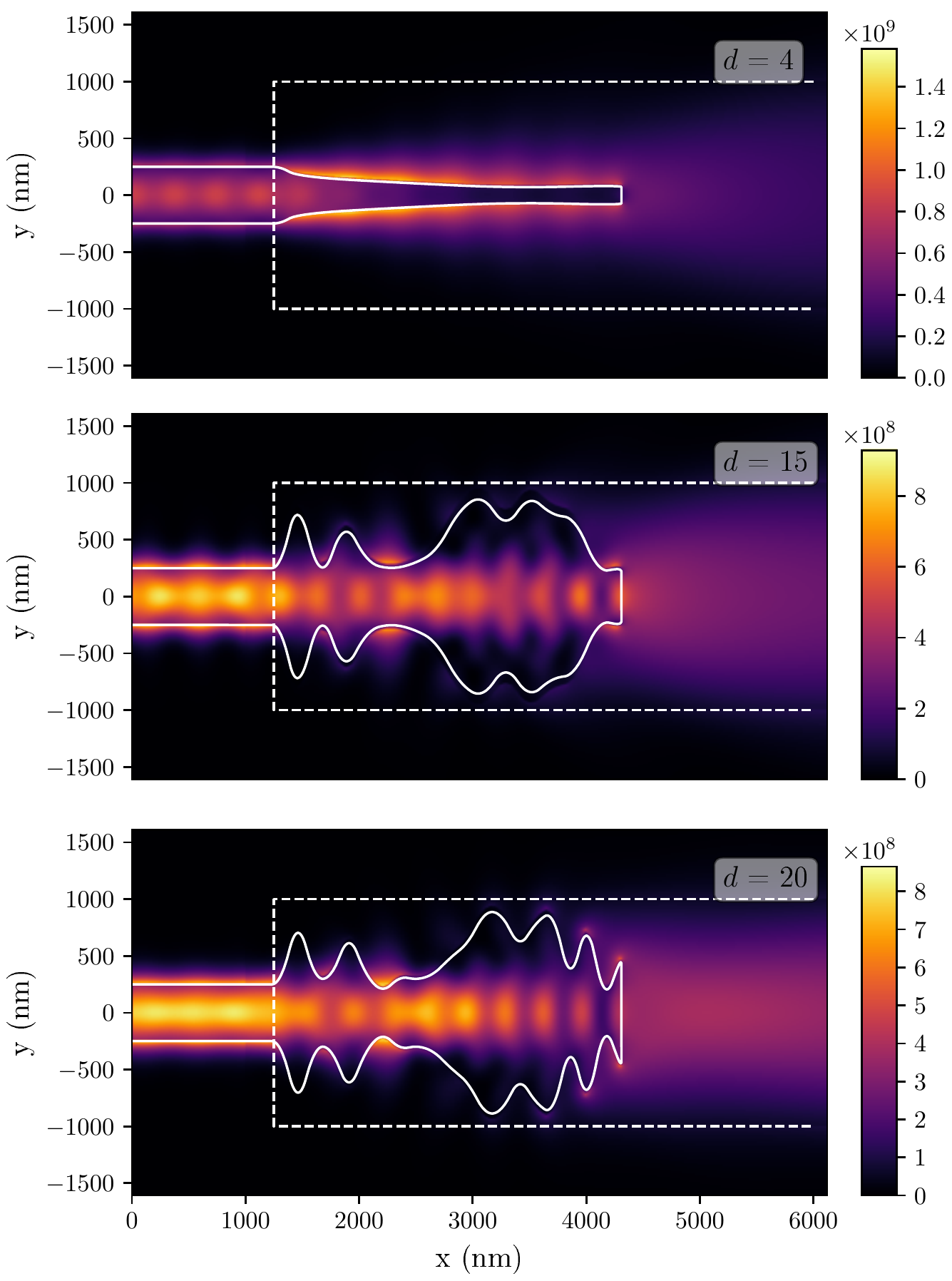}
\caption{Geometries of the obtained optimal waveguide couplers. The results show the optimal designs for couplers parametrized with a number $d$ of 4, 15, and 20 control points. The colors indicate the energy density of the electromagnetic field.}
\label{fig:optimal_geometries}
\end{figure}

Figure~\ref{fig:optimal_geometries} shows the optimal geometries obtained for the cases of $d$ equal to 4, 15, and 20. The optimal design for $d=4$ parameters has the shape of an adiabatic taper. The couplers parametrized with more control points present more complex geometries, although both the coupler obtained using 15 and 20 control points present similar characteristics between them. 

It is expected that these complex shapes are the result of maximizing the coupling efficiency at the target optimization wavelength of 1550 nm. But interesting enough, by taking a look at Fig.~\ref{fig:optimal_geometries}, one can see important differences between the field intensity for the designs where $d$ equals to 4 and 20 parameters. For the design with $d$ = 4, the width of the coupler is narrower than the silicon waveguide along the entire coupler region. As a result, the light is no longer confined within the silicon core, but it is radiated into the PWB waveguide section that has a lower refractive index. Since the width of the coupler decreases continuously and smoothly, the back-reflections along the coupler are small and they are produced continuously along the coupler. This can be concluded from the smooth profile of the field intensity within the core of the coupler. After 2 \micras{}, most of the incident light has already been radiated into the PWB waveguide section where it continues to be guided and the core carries almost no energy. Such a behaviour can be considered as the expected or rational solution to the problem.

The behavior of the coupler with $d$ = 20 is completely different. Its width is larger than the width of the silicon waveguide almost along the entire 3 \micras{} of the coupling section. As a consequence, the light is kept confined in the silicon core of the coupler and there is almost no radiation into the PWB waveguide until the terminating section of the coupler. The profile of the coupler presents a chirped subwavelength modulation. As a further difference to the design with $d$ = 4, an interference pattern is clearly present within the core section of the coupler. The strong interference is a clear indication that multiple reflections occur along the coupler. Nevertheless, the interferences between counter propagating waves in the input silicon waveguide are weaker than in the design for $d$ = 4. This observation is in agreement with the higher coupling efficiency shown in Fig.~\ref{fig:optimization_results}. The fact that the coupler with $d$ = 20 produces higher internal reflections but it presents a smaller total reflection at the input port can only be explained by destructive interference of the multiple reflections. The shape of the coupler seems to modulate the reflections in such a way that they interfere destructively at the design wavelength, allowing all the incoming power to be emitted into the PWB waveguide. At the same time, the shape at the region of the tip also modulates the radiation pattern, leading to an almost perfect overlap with the fundamental mode of the PWB. This alternative, completely non-adiabatic solution emerges somewhat unexpected.  

\begin{figure}[htb]
\centering
\includegraphics[width=\linewidth]{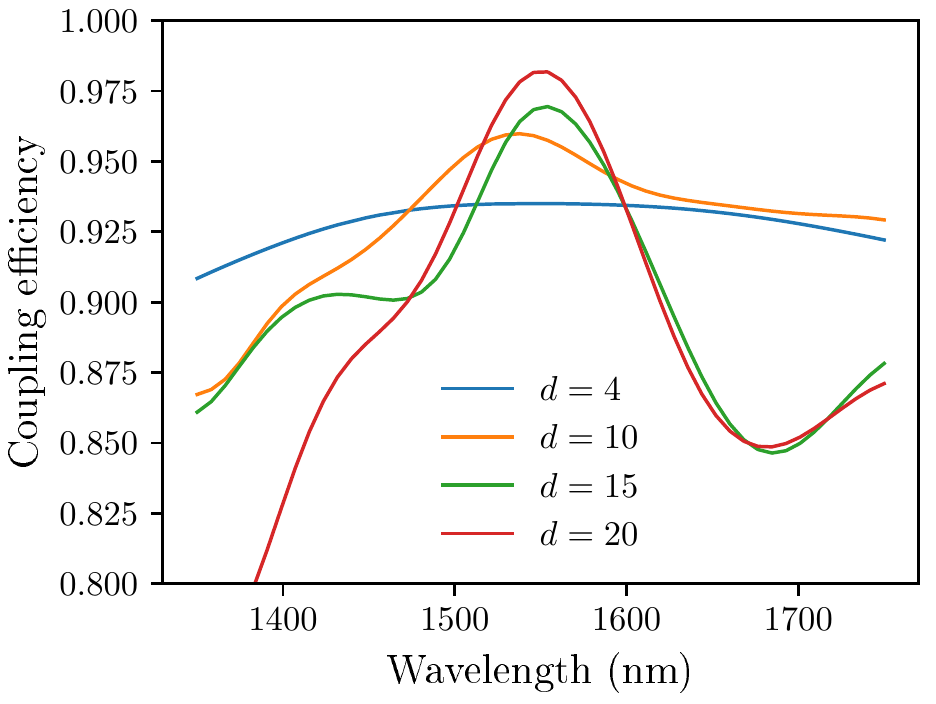}
\caption{Wavelength sweeps for the coupling efficiency of the optimal couplers.}
\label{fig:spectral_efficienty}
\end{figure}

To demonstrate how this fine tuning of the geometry affects the coupling efficiency in a spectral bandwidth around the central wavelength,  Fig.~\ref{fig:spectral_efficienty} shows a wavelength scan of the efficiencies of the optimized designs. The more complex couplers present a more pronounced, spectrally selective response. Their coupling efficiencies are higher than that of the adiabatic taper within a bandwidth of 100 nm. Beyond this spectral region the efficiencies drop below that of the adiabatic coupler, which maintains an efficiency above 0.9. 
 
\section{Conclusion}

We propose a global optimization scheme for the design of nanophotonic devices composed of a FEM solver with forward shape derivatives and an iterative Bayesian optimization algorithm. The iterative algorithm improves the main scalability problem of Bayesian optimization, allowing to access optimization problems that are not feasible for Bayesian optimization with a classical implementation, specially when derivative observations are included. With this approach, one can benefit from the good performance of Bayesian optimization for a wider range of devices. 

As an application example, we optimized the two dimensional model of a compact freeform waveguide coupler, achieving a coupling efficiency of 98 percent. The coupler is optimized for a target wavelength of 1550 nm and it can be implemented using e-beam lithography. The results show that a larger dimensional design space for the freeform shape leads to significantly higher coupling efficiencies at the target wavelength. On the other hand, the higher dimensional couplers are also more frequency selective.

\appendices
\section{Matrix update for $\matriz{L}_{K_{\rm ev}}$ and $\matriz{L}_{K_{\rm ev}}^{-1}$} \label{sec:appendix_matrix_update}

To derive the method to update $\matriz{L_{K_{\rm ev}}}$ based on a previously computed submatrix, we split the covariance matrix $\matriz{K_{\rm ev}}$ into four blocks,

\begin{equation}
\matriz{K_{\rm ev}} = \begin{bmatrix} \matriz{A} & \matriz{B}^T\\ \matriz{B} & \matriz{C} \end{bmatrix},
\end{equation}
where $\matriz{A}$ represents the covariance matrix computed in a subset of all evaluation points. In our case, $\matriz{A}$ will be the main block of $\matriz{K_{\rm ev}}$, and the submatrices $\matriz{B}$ and $\matriz{C}$ the result of adding a few more observations to the GP. However, this consideration about the relative sizes of the sub-blocks is not needed in the below derivation. 
It follows that $\matriz{L_{K_{\rm ev}}}$ can be obtained in terms of the elements of $\matriz{A}$, $\matriz{B}$ and $\matriz{C}$ as, Ref.~\cite{golub2012matrix} Sec. 6.5.4,

\begin{equation} \label{eq:cholesky_blocks}
\matriz{L_{K_{\rm ev}}} = \begin{bmatrix} \matriz{L_{A}} & \matriz{0} \\ \matriz{B} \cdot \matriz{L_{A}}^{-T} & \matriz{L_{S}} \end{bmatrix},
\end{equation}
with $\matriz{S}$ being the Schur complement of $\matriz{K_{\rm ev}}$,
\begin{equation}
\matriz{S} = \matriz{C}-\matriz{B}\cdot\matriz{A}^{-1}\cdot\matriz{B}^{T}.
\end{equation}

By defining the matrix $\matriz{X} = \matriz{B}\cdot\matriz{L_{A}}^{-T}$ one can express $\matriz{S}$ as 

\begin{equation}
\matriz{S} = \matriz{C}-\matriz{X}\cdot\matriz{X}^{T}.
\end{equation}

Here, we have used that the Cholesky decomposition of the inverse of a matrix equals to the inverse transpose of the Cholesky decomposition of the matrix. To see this, consider the Cholesky decomposition 
\begin{equation}\label{eq:Chol_M}\matriz{M} = \matriz{L_{M}}\cdot\matriz{L_{M}}^{T}
\end{equation}
of a matrix $\matriz{M}$.  By inverting Eq.~\eqref{eq:Chol_M} and comparing with the Cholesky decomposition of the inverse $\matriz{M}^{-1} = \matriz{L_{M^{-1}}}\cdot\matriz{L_{M^{-1}}}^{T}$ we have
\begin{equation}
\matriz{M}^{-1} = \matriz{L_{M^{-1}}}\cdot\matriz{L_{M^{-1}}}^{T} = \matriz{L_{M}}^{-T}\cdot\matriz{L_{M}}^{-1}. \label{eq:inverse_cholesky}
\end{equation}

To determine $\matriz{X}^T$ one can solve the triangular system $\matriz{L}_A \cdot \matriz{X}^T = \matriz{B}^T$.
Finally, we get the expression to the matrix update scheme,

\begin{equation}
\matriz{L_{K_{\rm ev}}} = \begin{bmatrix} \matriz{L_{A}} & \matriz{0} \\ \matriz{X} & \matriz{L_{S}} \end{bmatrix}.
\end{equation}

Equation~\eqref{eq:inverse_cholesky} shows how to efficiently use $\matriz{L_{K_{\rm ev}}}$ to compute the inference of $\sigma _p$, Eq.~\eqref{eq:inference_sigma}. If we define the vector $\ve{b}$ as

\begin{equation}
\ve{b} =  \matriz{L_{K_{\rm ev}}}^{-1} \cdot \matriz{K}_{*,{\rm ev}}^{T},
\end{equation}
we can infere the posterior $\sigma _p$ as
\begin{equation}
\matriz{K}_{*,{\rm ev}}\matriz{K}_{\rm ev}^{-1}\matriz{K}_{*,{\rm ev}}^{T} = \ve{b}^T \cdot \ve{b}.
\end{equation}

The vector $\ve{b}$ can be obtained solving the triangular system $\matriz{L_{K_{\rm ev}}}\cdot\ve{b} = \matriz{K}_{*,{\rm ev}}^{T}$ using forward and backward substitution.

One can also update the inverse of $\matriz{L_{K_{\rm ev}}}$ instead of $\matriz{L_{K_{\rm ev}}}$ itself,

\begin{equation}
\matriz{L_{K_{\rm ev}}}^{-1} =  \begin{bmatrix} \matriz{L_{A}}^{-1} & \matriz{0} \\ -\matriz{L_{S}}^{-1}\cdot\matriz{X}\cdot\matriz{L_{A}}^{-1} & \matriz{L_{S}}^{-1} \end{bmatrix}.
\end{equation}

Using the inverse of $\matriz{L_{K_{\rm ev}}}$ one can compute $\ve{b}$ as a matrix vector multiplication instead of calculating it with a forward and backward substitution, which halves the computation time of the inference. However, as we show in Appendix \ref{sec:appendix_stability}, working with $\matriz{L_{K_{\rm ev}}}^{-1}$ instead of $\matriz{L_{K_{\rm ev}}}$ leads to higher numerical errors as the covariance matrix $\matriz{K_{\rm ev}}$ is typically ill-conditioned.

\section{Stability analysis}\label{sec:appendix_stability}

It is well known, that the covariance matrix becomes ill-conditioned in the course of Bayesian optimization. Especially, when the observed sampling points are too close together, some rows of the covariance matrix are almost identical. The degeneracy of the matrix becomes even worse when using derivative information\cite{wu2017BOgradients}. It is therefore an important question if the presented iterative approach suffers from this degeneracy, i.e. if the GP inference is sufficiently accurate after a large number of iterations.

Many regularization techniques have been proposed\cite{GPregularization} in order to be able to work with an ill-conditioned covariance matrix. The most frequently used methods are the use of the LU decomposition (Ref.~\cite{gibbs1998bayesian} Section 3.1.1), the Cholesky decomposition (Ref.~\cite{GP_william} Section 2.2), and the use of the truncated singular value decomposition\cite{hansen1987truncatedsvd}, TSVD (see for example Ref.~\cite{gibbs1998bayesian} Section 3.1.2).
The LU decomposition requires 2/3 $\mathrm{N_{\rm ev}}^3$ floating point operations (flops) (Ref.~\cite{trefethen1997numerical} Lecture 23), the Cholesky decomposition requires 1/3 $\mathrm{N_{\rm ev}}^3$ flops (Ref.~\cite{trefethen1997numerical} Lecture 23), and the first phase of the SVD, which is in principle the most expensive part of the algorithm, requires 8/3 $\mathrm{N_{\rm ev}}^3$ flops when using the Golub-Kahan bidiagonalization (Ref.~\cite{trefethen1997numerical} Lecture 31). Note that, as discussed already in Appendix~\ref{sec:appendix_matrix_update}, twice the time is required to perform an inference when using the Cholesky decomposition in comparison with direct inversion methods, such as the SVD.

\begin{figure}[htb]
\centering
\includegraphics[width=\linewidth]{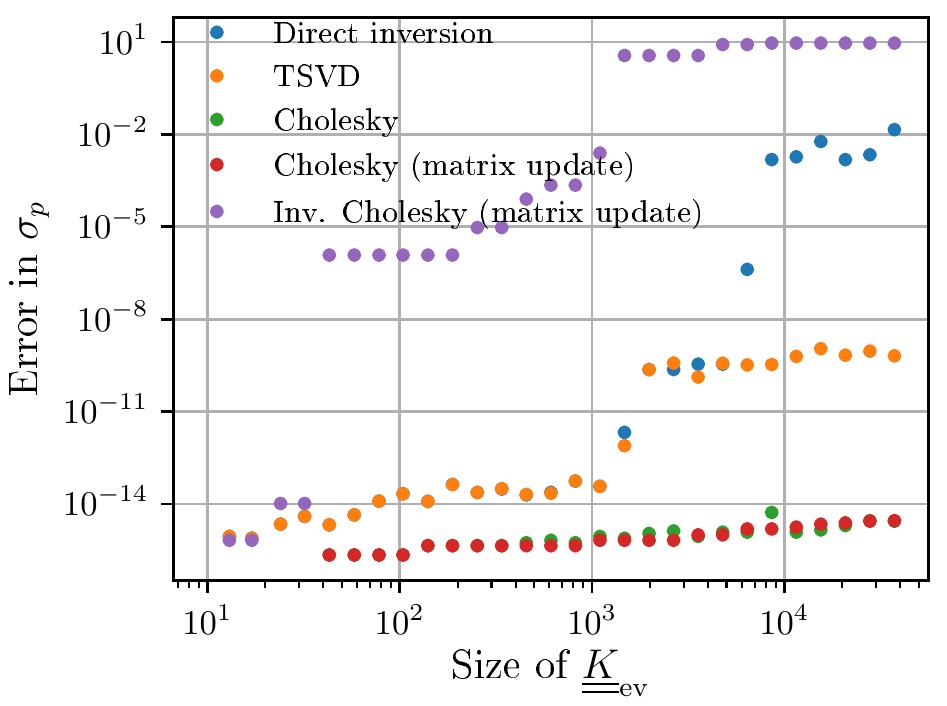}
\caption{Error produced by different numerical techniques used for solving the system of Eq.~ \eqref{eq:inference_sigma}: Direct inversion of $\matriz{K_{\rm ev}}$, the truncated singular value decomposition, the Cholesky decomposition on the two matrix updates approaches proposed in this work. The error shown is the maximum error obtained after evaluating $\sigma _p$ at all the evaluation points contained in the GP.}
\label{fig:regularization_test}
\end{figure}

Another noteworthy regularization technique that was proposed in the context of Bayesian optimization using derivative information is the spectral representation\cite{wu2017BOgradients} of the Gaussian process, SGP. The basic principle consists of Fourier transforming the design space and performing the inference in the frequency domain. By limiting the value for the highest frequency, the method can deal well with ill-conditioned covariance matrices.
However, the method is not suitable for the discussed applications, as it does not scale well with the number of dimensions $d$ of the design space. The method requires a fixed number $M$ of equidistant points in every dimension. The memory requirement for the transformed covariance matrix is of order $O(\mathrm{d^{2M}})$. Because of this memory requirement, we were not able to perform a comparison with this method. Note that in Ref.~\cite{wu2017BOgradients} the authors already mention that the method is memory efficient only when the length scales of the problem are large compared to the size of the design space.

To assess the numerical stability of the proposed matrix update schemes, we run a comparison of the error produced by the Cholesky decomposition and the TSVD decomposition and by both matrix update schemes. To measure the error, we infer the values of $\sigma _p$ at all the evaluation points contained in the GP after applying the different regularization techniques. The uncertainty should be ideally zero at the sampling points. The error is therefore defined as the maximum deviation from zero of $\sigma _p$ at all previous sampling points.
The results of the test are shown in Fig.~\ref{fig:regularization_test}. 

The behavior of both matrix update schemes is very different. While the matrix update scheme for the Cholesky decomposition keeps as stable as the Cholesky decomposition itself, the use of the inverse of the Cholesky decomposition produces large errors in $\sigma _p$ as the covariance matrix increases its size. Based on these results, the use of the inverse of the Cholesky decomposition seems to be not suitable to perform Bayesian optimization. However, the results also show us the high stability of the matrix update scheme for the Cholesky decomposition in comparison to the TSVD 
and the direct inversion. 
We attribute the stable behaviour to the fact that also algorithms for the full Cholesky decomposition compose the matrix row-by-row or column-by-column and do not require a pivoting of the matrix~\cite{golub2013matrix}. Therefore, it is possible to improve the scalability of Bayesian optimization without losing accuracy in the predictions of the GP.

\section*{Acknowledgements}
This work has received funding from the European Union's Horizon 2020 research and innovation programme under the
          Marie Sklodowska-Curie grant agreement No 675745.
          The authors also gratefully acknowledge financial support by the Deutsche Forschungsgemeinschaft (DFG)
          through – Project-ID 258734477 – SFB 1173 and – Project-ID 390761711 – EXC 2082/1.
          This project has received funding from the German Federal Ministry of Education and Research
          (BMBF Forschungscampus MODAL, project number 05M20ZBM) and by the Deutsche Forschungsgemeinschaft
          under Germany's Excellence Strategy - MATH+ (EXC-2046/1, project ID 390685689, AA4-6).
          X.G.S. acknowledges support from the Karlsruhe School of Optics and Photonics (KSOP).

\ifCLASSOPTIONcaptionsoff
  \newpage
\fi
\bibliographystyle{IEEEtran}
\bibliography{IEEEabrv,references_BO_with_derivatives}
\end{document}